# Scalable BFT Consensus Mechanism Through Aggregated Signature Gossip

Jieyi Long, Ribao Wei

*Theta Labs, Inc., San Jose, CA 95128, U.S.A*

{jieyi, ribao}@thetatoken.org

*Abstract*—In this paper, we present a new BFT consensus mechanism which enables thousands of nodes to participate in the consensus process, and supports very high transaction throughput. This is achieved via an aggregated signature gossip protocol which can significantly reduce the messaging complexity and thus allows a large number of nodes to reach consensus quickly. The proposed aggregated signature gossip protocol leverages a non-interactive leaderless BLS signature aggregation scheme. We have proven the correctness of the protocol, and analyzed its efficiency. In our analysis, each node only needs to send and receive $O(\log n)$ messages to reach agreement, where each message just contains a couple kilobytes of data.

## I. Introduction

For the blockchain technology to enter the mainstream, many technical challenges need to be tackled. These challenges include supporting high transaction throughput and fast confirmation time while maintaining a high level of decentralization.

Recent proposals like the Delegated Proof-of-Stake (DPoS) mechanism offers much higher throughput over Proof-of-Work chains like Bitcoin [1] and Ethereum [2]. However, they typically only permit a limited number of validator nodes [3, 4]. This sacrifices the level of decentralization and transaction security. Ideally, a Proof-of-Stake (PoS) based consensus mechanism should allow thousands of independent nodes, each with similar amounts of stake, to participate in the consensus process. To compromise such a system, an adversary needs to control a significant fraction of independent nodes, which is difficult to achieve.

In this paper we propose a novel **multi-level BFT consensus mechanism** that allows thousands of consensus participants, and yet can achieve high transaction throughput (1000+ TPS). The core idea is to have a small set of nodes, which forms the validator committee, to produce a chain of blocks as fast as possible using a PBFT-like process [5]. With a sufficient number of validators (e.g. 10 to 20), the committee can *produce* blocks at a fast speed, and it is already difficult enough for an adversary to compromise. Hence, it is reasonable to expect that they will produce a chain of blocks without forks with high probability. Then, all the thousand consensus participants, called the guardians, can finalize the chain generated by the validator committee. Here, finalization means to convince each honest guardian that more than 2/3 of all the other guardians possess the same chain of blocks.

Since there are many more guardians than validators, it could take longer time for the guardians to reach consensus than the validator committee. In order for the guardians to *finalize* the chain of blocks at the same speed as the validator committee produces new blocks, the guardian nodes can finalize the blocks at a coarser grain. To be more specific, they only need to agree on the hash of the checkpoint blocks, i.e. blocks whose heights are multiples of some integer $T$ (e.g. $T = 100$). This "leapfrogging" finalization strategy leverages the immutability characteristic of the blockchain data structure — as long as two guardian nodes agree on the hash of a block, with overwhelming probability, they will have exactly the same copy of the entire blockchain up to that block. Finalizing only the checkpoint blocks gives sufficient time for the thousands of guardians to reach consensus. Hence, with this strategy, the two independent processes, i.e., block production and finalization, can advance at the same pace.

Even though block finalization is only performed every $T$ blocks, we need to find a way to efficiently convince every honest guardian that indeed more than 2/3 of all guardians have the same checkpoint hash. A naive all-to-all broadcasting of the checkpoint block hash could work. However, it yields quadratic communication overhead, and so cannot scale to a large number of guardians. Instead we propose an **aggregated signature gossip** scheme which could significantly reduce the messaging complexity. The core idea is rather simple. Each guardian node keeps combining the partially aggregated signatures from all its neighbors, and then gossips out this aggregated signature, along with a compact bitmap which encodes the list of signers. This way the signature share of each node can reach other nodes with exponential speed thanks to the gossip protocol. Within $O(\log n)$ iterations, with high probability, all the honest guardian nodes should have aggregated the signatures from all other honest nodes if there is no network partition. Such fast convergence would allow us to choose a relatively small $T$ which translates into short block finalization time (in the order of minutes in practice).

The above is a high-level description of our proposal. We are actively implementing the proposed protocol in the Theta Ledger, a full-fledged blockchain designed for the video streaming industry. We have released our implementation on GitHub[1].

In the next section, we will first review the related works. Then, in Section III, we present the details and analysis of the multi-level BFT consensus protocol. Section IV provides experimental results which demonstrate the practicality of the proposed mechanism. Section V summarizes the paper with concluding remarks.

## II. Related Works

In recent years there have been many proposals that attempt to overcome the performance hurdle of the Proof-of-Work consensus. The Delegated Proof-of-Stake approach achieves high throughput, but at the expense of decentralization [3]. Algorand and DFINITY achieve a better level of decentralization by introducing a rotating set of committees selected at random to produce new blocks [6, 7]. However, both techniques lack a mechanism to allow other nodes to approve the block produced by the committee. Although unlikely, the random selection process could happen to elect a committee with more than two-third malicious

---

[1] https://github.com/thetatoken/theta-protocol-ledger

nodes. In such cases, the resulting blockchain might contain invalid blocks.

On a different track, there have been lines of work from Bitcoin-NG [13] to recent proposals like ByzCoin and OmniLedger [14, 16], that allow many nodes to jointly finalize new blocks. They employ a single leader to produce blocks in order to achieve high throughput, with a number of nodes validating the blocks afterwards. While bearing some similarities, in our approach we use a validator committee instead of a single leader, which significantly enhances security while maintaining high throughput. Furthermore, our guardian network leverages a BLS based signature aggregation scheme to reach agreement in a leaderless fashion. Compared to the Schnorr signature based CoSi scheme [15] employed by ByzCoin and OmniLedger, our approach is simpler to implement thanks to the non-interactive nature of the BLS signature aggregation. Moreover, the leaderless nature of our signature aggregation scheme could be much more robust in the presence of byzantine nodes compared to CoSi, which requires a leader to orchestrate all the nodes to produce the aggregate signatures. We will provide more in-depth comparison in the Remarks section.

We also note that the signature aggregation and the related threshold signature techniques have seen wide applications in blockchain research. For instance, they have been employed by researchers to construct random beacons, to achieve better censorship resilience, and to reduce storage costs [7, 18-20].

### III. THE MULTI-BFT CONSENSUS MECHANISM

#### A. Validators and Guardians

As mentioned in the introduction section, the validator committee is comprised of a limited set of validator nodes, typically in the range of ten to twenty. They can be selected through an EOS-like election process [3], or a randomized process. The committee may be subject to rotation to improve security, for instance using techniques like the "cryptographic sortition" method from Algorand [6], or the "threshold relay" mechanism from DFINITY [7]. To be eligible to join the validator committee, a node needs to lock up a certain amount of stake for a period of time. The locked stake could be slashed if malicious behavior is detected. The blocks that the committee reaches consensus on are called settled blocks, and the process by which they produce a chain of block is called the **block settlement process**.

The guardian network is a super set of the validator committee, i.e. a validator is also a guardian. The guardian network contains a large number of nodes, which could be in the range of thousands. With a certain amount of token lockup for a period of time, any node in the network can instantly become a guardian. The guardians download and examine the chain of blocks generated by the validator committee and try to reach consensus on the checkpoints. By allowing mass participation, we can greatly enhance transaction security. The blocks that the guardian network has reached consensus on are called finalized blocks, and the process by which they finalize the chain of blocks is called the **block finalization process**.

The name multi-level BFT consensus mechanism reflects the fact that the validator/guardian division provides multiple levels of security guarantee. The validator committee provides the first level of protection — with 10 to 20 validators, the committee can come to consensus at a fast pace. Yet it is resistant enough to attacks — in fact, it already provides a similar level of security compared to the DPoS mechanism. Thus, a transaction can already be considered safe when it has been included in a settled block, especially for low stake transactions, which are the most common transactions. The guardian network forms the second line of defense. With thousands of nodes, it is substantially more difficult for attackers to compromise, thus this provides a much higher level of security. A transaction is considered irreversible only when it is included in a finalized block. In the unlikely event that the validator committee is fully controlled by attackers, the guardians can re-elect the validators, and the blockchain can restart advancing from the latest block finalized by the guardians. To be more specific, the newly elected validator committee will start proposing and settling new blocks from the latest guardian-finalized block (those settled but not yet finalized blocks are discarded). And the guardians then follow the block finalization process described later in the paper to finalize the new blocks settled by the validator committee. We believe this mechanism can achieve a good balance among the level of decentralization, transaction throughput, and consistency, the three corners of the so-called "**impossible triangle**".

#### B. System Model

Before diving into the details of the block settlement and finalization process, we first list our assumptions of the system. For ease of discussion, below we assume each node (be it a validator or a guardian) has the same amount of stake. Extending the algorithms to the general case where different nodes have different amounts of stake is straightforward.

**Validator committee failure model**: There are $m$ validator nodes in total. Most of the time, at most 1/3 of them are byzantine nodes. They might be fully controlled by attackers, but we assume this happens only rarely. We also assume that between any pair of validators there is a direct message channel (e.g. a direct TCP socket connection).

**Guardian network failure model**: There are $n$ guardian nodes in total. At any moment, at most 1/3 of them are byzantine nodes. We do not assume a direct message channel between any two guardians. Messages between them might need to be routed through other nodes, some of which could be byzantine nodes.

**Timing model**: We assume the "weak synchrony" model. To be more specific, the network can be asynchronous, or even partitioned for a bounded period of time. Between the asynchronous periods there is a sufficient amount of time where all message transmissions between two directly connected honest nodes arrive within a known time bound. As we will discuss later in the paper, during the asynchronous period, the system simply stops finalizing new blocks. It would never produce conflicting blocks even during asynchrony periods. During the synchronous phase, block production will naturally resume, and eventual liveness can be achieved.

**Attacker model**: We assume powerful attackers. They can corrupt a large number of targeted nodes, but no more than 1/3 of all the guardians simultaneously. They can manipulate the network at a large scale, and can even partition the network but only for a bounded period of time. However, they are computationally bounded. For instance, they cannot forge fake signatures, and cannot invert cryptographic hashes.

*C. The Block Settlement Process*

Block settlement is the process by which the validator committee reaches agreement and produces a chain of blocks for the guardian network to finalize. Various PBFT based blockchain consensus algorithms including the EOS DPoS-BFT, Tendermint, Casper FFG, Hot-Stuff, Algorand, and DFINITY [3, 4, 6-9] can be employed for the validator committee to produce and settle on a chain of blocks. We will omit the details of these algorithm here. But it is worth pointing out that these algorithms guarantee safety when less than 1/3 of the validators have byzantine failures. Liveness can be achieved during the synchronous periods. When the network is partitioned, the validator committee stops generating new blocks, which is desirable since it avoids split-brain. However, after the network recovers, the validator committee can continue to produce new blocks. With ten to twenty validators, the committee can produce blocks with fast speed, enabling 1000+ transactions per second throughput. Meanwhile, if these validators are run by different entities, it is already reasonably difficult for an attacker to gain full control of the committee. Thus, the guardians can expect that the validator committee can produce a chain of blocks without forks most of the time.

*D. The Block Finalization Process*

As the validator committee adds new blocks to the blockchain, the guardian network finalizes the produced chain in parallel. As mentioned earlier, to finalize the chain of blocks generated by the validator committee, the guardians only need to reach consensus on the hashes of the **checkpoint blocks**, which are the blocks whose heights are multiple of some integer $T$ (e.g. $T = 100$).

To see why it is sufficient to finalize just the checkpoint blocks, we note that the transaction execution engine of the blockchain software can be viewed as a "deterministic state machine", whereas a transaction can be viewed as a deterministic state transfer function. If two nodes run the same state machine, then from an identical initial state, after executing the same sequence of transactions, they will reach an identical end state. Note that this is true even when some of the transactions are invalid, as long as those transactions can be detected by the state machine and skipped. For example, assume a transaction tries to spend more tokens than the balance of the source account. The state machine can simply skip this transaction after performing the sanity check. This way the "bad" transactions have no impact on the final state.

In the context of blockchain, if all the honest nodes have the same copy of the blockchain, they can be ensured to arrive at the same end state after processing all the blocks in order. But with one caveat — the blockchain might contain a huge amount of data. How can two honest nodes compare whether they have the same chain of blocks efficiently?

Here the immutability characteristic of the blockchain data structure becomes highly relevant. Since the header of each block contains the hash of the previous block, as long as two nodes have the same hash of a checkpoint block, with overwhelming probability, they should have an identical chain of blocks from the genesis up to that checkpoint. Of course each of the guardian nodes needs to verify the integrity of the blockchain. In particular, they need to verify that the block hash embedded in each block header is actually the hash of the previous block. We note that a node can perform the integrity check on its own; *no communication* with other nodes is required.

Interestingly, the immutability characteristic also enhances the tolerance to network asynchrony or even partition. With network partition, the guardians may not be able to reach consensus on the hash of a checkpoint. However, after the network is recovered, they can move on to vote on the next checkpoint. If they can then reach agreement, all the blocks up to the next checkpoint are considered finalized, regardless of whether or not they have consensus on the prior checkpoints.

To provide byzantine fault tolerance, an honest node needs to be assured that more than 2/3 of the guardians has the same checkpoint block hash. Hence it needs to receive signatures for a checkpoint hash from more than 2/3 of all guardians before the node can mark that checkpoint as finalized. This is to ensure safety, which is similar to the "commit" step in the PBFT protocol.

Since the guardians only need to vote on checkpoint hashes every $T$ blocks, they have more time to reach consensus. A straightforward implementation of checkpoint finalization is thus to follow the PBFT "commit" step where each guardian broadcasts its signature of the checkpoint hash to all other guardians. This requires each node to send, receive and process $O(n)$ messages, where each message can be a couple kilobytes long. Even with $T$ blocks time, this approach still cannot scale beyond a couple hundred guardian nodes, unless we select a large $T$ value, which is undesirable since it increases the block finalization latency.

*E. Scale to Thousands of Guardian Nodes*

To reduce the communication complexity in order to scale to thousands of guardian nodes, we have designed an **aggregated signature gossip protocol** inspired by the BLS signature aggregation scheme [10, 11] and the gossip protocol. The protocol requires each guardian node to process a much smaller number of messages to reach consensus. Below are the steps of the aggregated signature gossip protocol. It uses the BLS algorithm for signature aggregation.

The core idea is rather simple. Each guardian node keeps combining the partially aggregated signatures from its neighbors, and then gossips out the newly aggregated signature. This way the signature share of each node can reach other nodes at an exponential speed thanks to the gossip protocol. On the other hand, the signature aggregation keeps the size of the messages small.

Algorithm 1 provides the pseudo code of the protocol. Here $i$ is the index of the current guardian node. The first line of the protocol uses function **SignBLS**() to generate its initial aggregated signature $\sigma_i$. It essentially signs a message which is the concatenation of the height and hash of the checkpoint block using the BLS signature algorithm:

$$h_i \leftarrow H(pk_i, height_{cp} || hash_{cp}) \quad (1)$$

$$\sigma_i \leftarrow h_i^{sk_i} \quad (2)$$

In the first formula above, function $H : G \times \{0,1\}^* \to G$ is a hash function that takes both the public key $pk_i$ and the message as input. This is to prevent the rogue public-key attack [12]. Here $G$ is a multiplicative cyclic group of prime order $p$ with generator $g$.

**Protocol: Aggregated Signature Gossip**

```
1   finalized ← false, σ_i ← SignBLS(sk_i, height_cp || hash_cp), ĉ_i ← InitSignerVector(i)
2   for t = 1 to L begin
3       send (σ_i, ĉ_i) to all its neighboring guardians
4       if finalized break
5       wait for (σ_j, ĉ_j) from all neighbors until timeout
6       verify each (σ_j, ĉ_j), discard if it is invalid
7       aggregate valid signatures σ_i ← σ_i · ∏_j σ_j, ĉ_i ← (ĉ_i + ∑_j ĉ_j) mod p
8       calculate the number of unique signers s ← ∑_{u=1}^{n} I(c_{iu} > 0)
9       if s ≥ 2n/3 finalized ← true
10  end
```

Algorithm 1. The aggregated signature gossip protocol

The protocol also uses function **InitSignerVector()** to initialize the signer vector $\hat{c}_i$, which is an $n$ dimensional integer vector whose $u^{th}$ entry represents how many times the $u^{th}$ guardian has signed the aggregated signature. After initialization, its $i^{th}$ entry is set to 1, and the remaining entries are all set to 0.

After the initialization, the guardian enters a loop. In each iteration, the guardian first sends out its current aggregated signature $\sigma_i$ and signer vector $\hat{c}_i$ to all its neighbors. Then, if it has not considered the checkpoint as finalized, it waits for the signatures and signer vectors from all its neighbors, or waits until timeout. Upon receiving all the signature and signer vectors, it checks the validity of $(\sigma_j, \hat{c}_j)$ received from neighboring node $j$ using the BLS aggregated signature verification algorithm.

$$h_u \leftarrow H(pk_u, height_{cp} || hash_{cp}) \quad (3)$$

$$\text{check if } e(\sigma_j, g) = \prod_{u}^{n} (e(h_u, pk_u))^{c_{ju}} \quad (4)$$

where $c_{ju}$ is the $u^{th}$ entry of vector $\hat{c}_j$, and $e : G \times G \to G_T$ is a *bilinear mapping* function from $G \times G$ to $G_T$, another multiplicative cyclic group also of prime order $p$. To guard against byzantine nodes, all the invalid signatures and their associated signer vectors are discarded for the next aggregation step. It is worth pointing out that besides $height_{cp}$ and $hash_{cp}$, the above check also requires the public key $pk_u$ of the relevant guardians as input. All these information should be available locally, since when a guardian locks up its stakes, its public key should be attached to the stake locking transaction which is already written into the blockchain. Hence, no communication with other nodes is necessary to retrieve these inputs.

The aggregation step aggregates the BLS signature $\sigma_j$, and updates the signer vector $\hat{c}_j$ from the neighbors. Note that for the vector update, we take $\bmod\, p$ for each entry. We can do this because $e(h_u, pk_u) \in G_T$, which is a multiplicative cyclic group of prime order $p$. This guarantees that vector $\hat{c}_j$ can always be represented with a bounded number of bits.

$$\sigma_i \leftarrow \sigma_i \cdot \prod_j \sigma_j, \quad \hat{c}_i \leftarrow (\hat{c}_i + \sum_j \hat{c}_j) \bmod p \quad (5)$$

The algorithm then calculates the number of unique signers of the aggregated signature for node $i$.

$$s \leftarrow \sum_{u=1}^{n} I(c_{iu} > 0) \quad (6)$$

Here $c_{iu}$ is the $u^{th}$ entry of vector $\hat{c}_i$. Function $I: \{\text{true, false}\} \to \{1, 0\}$ maps a true condition to 1, and false to 0. Condition $c_{iu} > 0$ indicates that node $u$ has signed the aggregated signature that node $i$ currently possesses at least once. Hence the summation counts how many unique signers have contributed to the aggregated signature. If the signature is signed by more than 2/3 of all the guardians, the guardian can consider the checkpoint as finalized.

If the checkpoint is finalized, the aggregated signature will be gossiped out in the next iteration. After this gossip, within $O(\log(n))$ iterations all the honest guardians will have an aggregated signature signed by more than 2/3 of all the guardians if the network is not partitioned. The for-loop has $L$ iterations, $L$ should be in the order of $O(\log(n))$ to allow the signature to propagate through the network.

*F. Analysis*

*1) Finalization Safety and Liveness Analysis*

*a) Safety*

Safety of the block finalization is easy to prove. Under the 2/3 supermajority honesty assumption, if two checkpoint hashes for the same height both have signatures from more than 2/3 of the guardians, then by the pigeonhole principle, more than 1/3 of the guardians must have signed both. Given that more than 2/3 of the guardians are honest, this implies at least one honest guardian signed both, which implies they are the same checkpoint hash.

*b) Liveness*

Without network partition, as long as $L$ is large enough, it is highly likely that after $O(\log(n))$ iteration, all the honest nodes will see an aggregated signature that combines the signatures of all honest signers. This is similar to how the gossip protocol can robustly spread a message throughout the network in $O(\log(n))$ time, even with up to 1/3 byzantine nodes. When there is network partition, consensus for a checkpoint may not be able to be reached. However, after the network partition is over, the guardian network can proceed to finalize the next checkpoint block. If consensus can then be reached, all the blocks up to the next checkpoint are considered finalized. Hence the finalization process will progress eventually.

Note that the above analysis only covers the safety and liveness property of the block finalization process. The analysis for the block settle process is omitted in this paper due to page limit. Please refer to the Theta Technical Whitepaper for the full analysis [17].

*2) Aggregated Signature Gossip Protocol Analysis*

*a) Correctness of the Protocol*

To prove the correctness of the aggregated signature gossip protocol, we need to prove two claims. First, if an aggregated signature is correctly formed by honest nodes according to Algorithm 1, it can pass the check given by Formula (4). Second, the aggregated signature is secure against forgery.

To prove **the first claim**, we note that both group $G$ and $G_T$ are cyclic groups. A cyclic group is always an abelian group, and hence we can switch the order of elements in multiplications. Furthermore, since group $G_T$ has a prime order $p$, given a group element $v \in G_T$, for any integer $c$, $v^c = v^{c \bmod p}$. With these properties, it is straightforward to prove the first claim using the definition of bilinear mapping. In the proof below, $Nb$ denotes the set of neighboring nodes of node $j$, $\sigma_{jk}$ is the partially aggregated signature collected from neighbor node $k$ in the current iteration, and $u$ is the signature of node $u$ given by Formula (2).

$$e(\sigma_j, g) = e\left(\prod_{k \in Nb} \sigma_{jk}, g\right) = \prod_{k \in Nb} e(\sigma_{jk}, g)$$
$$= \prod_{u=1}^{n}\left(e(\sigma_u, g)\right)^{c_{ju}} = \prod_{u=1}^{n}\left(e\left(h_u^{sk_u}, g\right)\right)^{c_{ju}}$$
$$= \prod_{u=1}^{n}\left(e(h_u, g^{sk_u})\right)^{c_{ju}} = \prod_{u=1}^{n}\left(e(h_u, pk_u)\right)^{c_{ju}}$$

Now let us prove **the second claim** regarding the unforgeability of the aggregated signature. To prove this claim, we will show that signature forgery can be reduced to the Computational Diffie-Hellman (CDH) problem [10, 11]. Stated more formally, we will prove that for randomly chosen $pk_1 = g^{sk_1}, \ldots, pk_n = g^{sk_n} \in G$, and random message hashes $h_1, \ldots, h_n \in G$, finding $\sigma \in G$ and integers $c_1, c_2, \ldots, c_n$ which satisfy the equation below is as hard as solving the CDH problem.

$$e(\sigma, g) = \prod_{u=1}^{n}\left(e(h_u, pk_u)\right)^{c_u}$$

Let us first look at a special case where $n = 1$, i.e., there is only one guardian node. In this case, for the adversary, forging a single guardian's aggregated signature means to find $\sigma \in G$ and an integer $c$ that satisfies

$$e(\sigma, g) = \left(e(h, pk)\right)^c \qquad (7)$$

Note that
$$\left(e(h, pk)\right)^c = e(h^c, pk) = e(h^c, g^{sk}) = e(h^{sk \cdot c}, g)$$

We thus have
$$\sigma = (h^{sk})^c$$

Next, using the Extended Euclidean algorithm, in polynomial time we can find an integer $d$ such that $c \cdot d \equiv 1 \bmod p$, where $p$ is the order of $G$, which is a prime number. Thus

$$\sigma^d = (h^{sk})^{c \cdot d} = (h^{sk})^{c \cdot d \bmod p} = h^{sk}$$

which means that the adversary can compute $h^{sk}$ from $g$, $g^{sk}$, and $h$ in polynomial time using $\sigma$ and $c$. This indicates that finding $\sigma \in G$ and integer $c$ satisfying equation (7) is as hard as the CDH problem.

When there are $n > 1$ guardian nodes, signature forgery means that the adversary must find $\sigma \in G$, and integers $c_1, c_2, \ldots, c_n$ that satisfies the equation below

$$e(\sigma, g) = \prod_{u=1}^{n}\left(e(h_u, pk_u)\right)^{c_u} \qquad (8)$$

Again, we show this is equivalent to solving the CDH problem. To see why, for randomly chosen $g$, $g^{sk_1}$ and $h_1$, the adversary can randomly generate $n - 1$ key pairs $(sk_2, pk_2), \ldots (sk_n, pk_n)$, where $pk_u = g^{sk_u}$ for $u = 2, 3, \ldots, n$. He then plugs these values into Equation (8), and solves for $\sigma \in G$ and integers $c_1, c_2, \ldots, c_n$. Note that

$$e(\sigma, g) = \prod_{u=1}^{n}\left(e(h_u, pk_u)\right)^{c_u}$$
$$= \prod_{u=1}^{n} e\left(\left(h_u^{sk_u}\right)^{c_u}, g\right)$$
$$= e\left(\prod_{u=1}^{n}\left(h_u^{sk_u}\right)^{c_u}, g\right)$$

This means

$$\sigma = \prod_{u=1}^{n}\left(h_u^{sk_u}\right)^{c_u} \qquad (9)$$

Since the adversary knows $sk_2, \ldots, sk_n$, he can plug these private keys into Equation (9), which then gives

$$\left(h_1^{sk_1}\right)^{c_1} = \sigma \cdot \left(\prod_{u=2}^{n}\left(h_u^{sk_u}\right)^{c_u}\right)^{-1}$$
$$= \sigma \cdot \prod_{u=2}^{n} h_u^{-sk_u \cdot c_u}$$

Similar to the reasoning for $n = 1$, in polynomial time we can find an integer $d_1$ such that $c_1 \cdot d_1 \equiv 1 \bmod p$. Thus we have

$$\left(\sigma \cdot \prod_{u=2}^{n} h_u^{-sk_u \cdot c_u}\right)^{d_1} = \left(h_1^{sk_1}\right)^{c_1 \cdot d_1}$$
$$= \left(h_1^{sk_1}\right)^{c_1 \cdot d_1 \bmod p} = h_1^{sk_1}$$

Hence, after solving Equation (8), the adversary can efficiently compute $h_1^{sk_1}$ for randomly chosen $g$, $g^{sk_1}$ and $h_1$. This implies that solving Equation (8) is as hard as solving the CDH problem.

In summary, we can conclude that our proposed aggregated signature gossip protocol is correct and secure.

*b) Computational Complexity*

The aggregated signature gossip algorithm requires each guardian node to validate the correctness of the combined signatures from its neighbors (Line 6 in Algorithm 1) using Formula (4). At first glance, it appears that for each signature validation, the node needs to perform $O(n)$ pairing calculations, i.e., $e(h_u, pk_u)$ for $u = 1, 2, \ldots, n$, where $n$ is the number of guardian nodes. However, we note that these pairing results are reusable for other signatures for the same checkpoint. Thus, only $O(n)$ pairing calculations are necessary for finalizing each checkpoint block. These pairing results can even be pre-computed since both $h_u$ and $pk_u$ can both be derived from the blockchain data available locally. Then, for each signature validation using Formula (4), the node only needs to raise each $e(h_u, pk_u)$ to a relatively small power $c_u$ (see the next section on the analysis on $c_u$), and then multiply the results together. On a modern processor, a pairing computation can be done within a couple of milliseconds. Thus, computing the required pairings for thousands of nodes takes only a few seconds. Hence, signature validation is not a bottleneck for block finalization, as the finalization time is designed to be in the order of minutes.

*c) Messaging Complexity*

The aggregated signature gossip protocol runs for $L$ iterations, which is in the order of $O(\log(n))$, where $n$ is the number of guardian nodes. In each iteration, the guardian needs to send message $(\sigma_i, \hat{c}_i)$ to all its neighbor nodes. Depending on the network topology, typically it is reasonable to assume that for an average node, the number of neighboring nodes is a constant (i.e. the number of neighbors does not grow as the total number of nodes grows). Hence the number of messages a node needs to send/receive to finalize a checkpoint is in the order of $O(\log(n))$, which is much better than the $O(n)$ complexity in the naive all-to-all signature broadcasting implementation. We do acknowledge that each message between two neighboring guardians contains an $n$ dimensional signer vector $\hat{c}_i$, where each entry of $\hat{c}_i$ is an integer smaller than prime $p$. Below we will present a theoretical model for the messaging complexity, which shows that this vector can be represented rather compactly since most of its entries are small integers ($\ll p$).

To model the message size, we would need to introduce the following notations. First, we define vector $\hat{x}_{k,t}$, which is composed of the $k^{th}$ element of the signer vector $\hat{c}_j$ that each guardian node possesses in iteration $t$

$$\hat{x}_{k,t} = (c_{1k,t}, c_{2k,t}, \ldots, c_{nk,t})^T$$

As a special case, initially when $t = 0$, $\hat{x}_{k,0} = (0, 0, \ldots, 1, \ldots, 0)^T$ where only the $k^{th}$ element is 1. We also define matrix $\boldsymbol{\Omega}$ which represents the connectivity among the guardian nodes.

$$\boldsymbol{\Omega} = (\delta_{ij})_{n \times n}$$

where $\delta_{ij} = 1$ only when $i = j$, or node $i$ and node $j$ are neighbors (i.e., there is a direct TCP socket connection between node $i$ and node $j$). Otherwise $\delta_{ij} = 0$.

It can be shown that in Algorithm 1, if we do not break the loop (i.e. remove Line 4), the following equation holds

$$\hat{x}_{k,t} = \boldsymbol{\Omega} \cdot \hat{x}_{k,t-1}$$

Thus we have

$$\hat{x}_{k,t} = \boldsymbol{\Omega}^t \cdot \hat{x}_{k,0}$$

We know that if $\det(\boldsymbol{\Omega}) \neq 0$, matrix $\boldsymbol{\Omega}$ always has an eigendecomposition

$$\boldsymbol{\Omega} = \boldsymbol{V}(\lambda_i)_{n \times n} \boldsymbol{V}^{-1}$$

where $(\lambda_i)_{n \times n}$ is an $n \times n$ matrix whose diagonal are the eigenvalues, and the non-diagonal elements are all zero. Hence we have

$$\hat{x}_{k,t} = \boldsymbol{V}(\lambda_i)_{n \times n}^t \boldsymbol{V}^{-1} \cdot \hat{x}_{k,0}$$

This indicates that the biggest entry of $\hat{x}_{k,t}$, and hence the signer vectors, is in the order of $\lambda_{max}^t$, where $\lambda_{max}$ is the biggest eigenvalue of the connectivity matrix. Although at first glance, the entries increase exponentially, we note that Algorithm 1 runs only for $O(\log n)$ iterations. Thus it can be shown that the biggest entries is in the order of $O(n)$, which can be represented with $O(\log n)$ bits.

Notice that in the above analysis, we assumed Line 4 in Algorithm 1 is removed. With Line 4, a guardian node stops aggregating signature and signer vectors after its current signature contains shares from more than 2/3 of all guardians. This can further reduce the bits required to represent the signer vectors.

To get a more concrete idea of the messaging complexity, let us work out an example. Assume that we pick a 170-bit long prime number $p$ for the BLS signature, which can provide security comparable to that of a 1024-bit RSA signature. And there are 1000 guardian nodes in total. Under this setting, $\hat{c}_i$ can be naively represented with about twenty kilobytes without any compression. However, since most of the entries of $\hat{c}_i$ are in the order of $O(n)$ (i.e. roughly 1000), which is far smaller than $p$, $\hat{c}_i$ can be compressed very effectively to a couple kilobytes long. Plus the aggregated signature, the size of each message is typically in the kilobytes range. Moreover, if we assume on average a guardian connects to 20 other guardians, then $L$ can be as small as 5 (more than twice of $\log_{20}(1000) = 2.3$). This means finalizing one checkpoint just requires a guardian to send/receive around 100 messages to/from its neighbors, each about a couple kilobytes long. The next Experimental Results section will provide more data points.

IV. EXPERIMENTAL RESULTS

We have implemented a simulator in Python to simulate the behavior of the guardian network, and have conducted various experiments to evaluate the messaging complexity of the checkpoint block finalization process. To setup the simulation, we note that in a production blockchain, typically when a guardian node joins the network, it first connects to a few seed nodes. These seed nodes then send a list of candidate peer nodes to the new node, and the new node randomly selects a subset to connect to. This process continues until the maximum neighbor count is reached. In our experiments, we generated random graphs to simulate the connections among the guardian nodes following this process.

We have simulated guardian networks with 1000, 2000, and 3000 nodes under different settings with byzantine node ratio ranges from 0% to 30%. These byzantine nodes either send their neighbors fake signatures, or send out no signature.

Table I and Table II summarize the experimental results. Table I is for the scenario where each guardian node has 20 neighboring nodes on average. The simulation results for the 30 neighboring nodes case are provided in Table II.

TABLE I. MESSAGING COMPLEXITY SIMULATION (AVG(|NB|) = 20)

| $n$ | Setting 1 | | Setting 2 | | Setting 3 | | Setting 4 | |
|---|---|---|---|---|---|---|---|---|
| | byz | max entry | byz | max entry | byz | max entry | byz | max entry |
| 1000 | 0% | 125 | 10% | 116 | 20% | 103 | 30% | 473 |
| 2000 | 0% | 112 | 10% | 104 | 20% | 445 | 30% | 1053 |
| 3000 | 0% | 108 | 10% | 453 | 20% | 817 | 30% | 1278 |

TABLE II. MESSAGING COMPLEXITY SIMULATION (AVG(|NB|) = 30)

| $n$ | Setting 1 | | Setting 2 | | Setting 3 | | Setting 4 | |
|---|---|---|---|---|---|---|---|---|
| | byz | max entry | byz | max entry | byz | max entry | byz | max entry |
| 1000 | 0% | 161 | 10% | 166 | 20% | 153 | 30% | 139 |
| 2000 | 0% | 161 | 10% | 143 | 20% | 135 | 30% | 376 |
| 3000 | 0% | 146 | 10% | 142 | 20% | 132 | 30% | 675 |

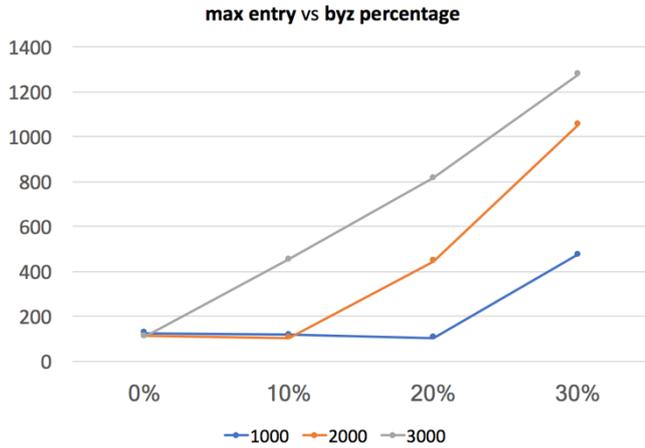

Figure 1. Messaging complexity simulation results for case where Avg(|Nb|) = 20

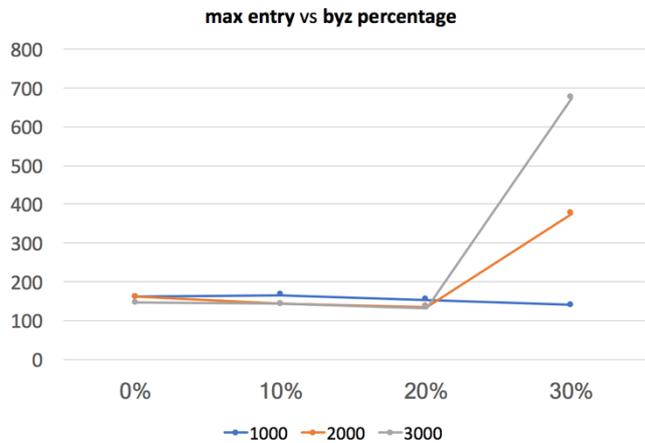

Figure 2. Messaging complexity simulation results for case where Avg(|Nb|) = 30

The column *n* represents the total number of guardian nodes, column *byz* is the percentage of byzantine nodes, and column *max entry* provides the biggest entry of $\hat{c}_i$ among **all** guardian nodes upon convergence. Here convergence means each honest guardian node receives an aggregated signature with shares from more than 2/3 of all nodes. In other words, the checkpoint is finalized. Since the messages among the guardian nodes are dominated by $\hat{c}_i$, the biggest entry of $\hat{c}_i$ determines the size of the messages.

These results are also visualized in Figure 1 and Figure 2, where the y-axis represents the *max entry*, and the x-axis is the percentage of byzantine nodes. The three lines in different colors plot the trends of the *max entry* versus *byzantine percentage* for 1000, 2000, and 3000 guardian nodes, respectively.

From the simulation result, under most settings, the biggest entry of $\hat{c}_i$ of all guardian nodes generally increases as the percentage of byzantine nodes becomes larger. However, in most cases, it is less than 1024. This means that each entry of the signer vector $\hat{c}_i$ can be represented by 10 bits. In many cases, the biggest entry is less than 256, and thus can be represented with a single byte. In any case, even with a couple thousand guardian nodes, the size of each messages is just a couple kilobytes (the number of entries in a signer vector is the same as the number of guardian nodes).

Finally, we observed that the simulation converges very quickly. In all of our simulations, the process terminated within 5 iterations. This is in the order of $O(\log_b n)$, where $b$ is the average neighboring nodes of each guardian. Since each guardian has 20 to 30 neighbors, this means each honest guardian node only needs to send/receive 100 to 150 messages in total to reach consensus for a checkpoint block, even when there are 30% byzantine nodes in the network. Thus, each checkpoint block finalization would only take a couple of minutes to complete. This indicates our proposed aggregated signature gossip protocol is very practical and robust even with a high percentage byzantine nodes.

V. REMARKS

Firstly, we would like to point out that the multi-level BFT frameworks can be applied for both permissioned blockchains and permissionless blockchains. In the permissioned chain setting, each node (be it a validator or guardian) may have the same voting power, and block finalization requires signatures from more than 2/3 of all guardian nodes. Also as mentioned earlier, techniques like "cryptographic sortion" and "threshold relay" from Algorand and DFINITY can be borrowed for electing and rotating the validator committee. In the permissionless setting, any node can become a guardian by staking a certain amount of tokens above the required minimum. Voting power of a node is proportional to the amount of tokens it staked.

Next, we also would like to provide a more detailed comparison with ByzCoin and OmniLedger.

1) We note that our aggregated signature gossip protocol is more robust than ByzCoin and OmniLedger in the presence of byzantine nodes. ByzCoin uses a single leader (i.e. the current block producer) to form the communication tree. With a tree topology, the root leader could become a single point of failure. Even when the leader is not faulty, if an intermediate node close to the root node is byzantine, the leader might not be able to collect sufficient number of signature shares. In such cases, ByzCoin falls back to the direct communication approach (i.e. the star topology), where the leader has to take $O(n)$ time to ask each node for their signature. OmniLedger proposed an improvement to ByzCoinX which uses a two-level tree to improve the tolerance to byzantine nodes [16]. Yet it still requires the "protocol leader" to coordinate the nodes into a communication tree — the protocol leader is in charge of selecting the group leaders, which are responsible for managing the communication with their own group members. If any group leader is a byzantine node, it can be replaced by the protocol leader. While being more robust to byzantine faults, this approach still seems to rely on the protocol leader, which could still be a single point of failure. Moreover, the two-level leader structure might add extra coordination and communication complexity to the protocol. In comparison, our gossip based protocol is leaderless. Even with a high number of byzantine nodes, the signature share of a node will propagate to most of the honest nodes in $O(\log n)$ time, similar to message gossiping.

2) Our aggregated signature gossip scheme is much simpler to implement. The CoSi protocol used by ByzCoin and OmniLedger is based on the Schnorr signature aggregation, which is an interactive protocol. In particular, in order to generate an aggregated Schnorr-signature for a message, all the singers need to collaborate to generate a shared challenge string (i.e. $c = H(\hat{V}_0||S)$ in the "Challenge" phase of the CoSi protocol [15]). The challenge string is required for each signer to generate its signature share. Hence, a leader is required to orchestrate the whole process, and the protocol needs to take multiple rounds. In comparison, our proposed protocol is based on the non-interactive BLS signature aggregation scheme, which eliminates the need for coordination among the nodes. Moreover, instead of the communication tree, we use the gossip protocol to propagate the aggregated signatures. These lead to a simple and robust design which could be easier to implement in practice.